\documentclass[fleqn,twoside]{article}
\usepackage{espcrc2}

\usepackage{graphicx}
\input epsf
\usepackage[figuresright]{rotating}

\newcommand{\AmS}{{\protect\the\textfont2
  A\kern-.1667em\lower.5ex\hbox{M}\kern-.125emS}}

\title{The Cyclic Universe: An Informal Introduction }

\author{Paul J. Steinhardt
\address[MCSD]{Joseph Henry Laboratories, 
        Department of Physics,\\
        Princeton University, Princeton, NJ 08544, USA.}
 Neil Turok\address{
DAMTP, Centre for Mathematical Sciences,
	  Wilberforce Road, Cambridge, CB3 0WA, UK.}}
\begin{document}

\begin{abstract}
The Cyclic Model is a radical,
new cosmological scenario which proposes that the Universe
undergoes
an endless sequence
of  epochs which begin with  a `big bang'
and end in a `big crunch.'  When the Universes bounces
from contraction to re-expansion, the temperature and density
remain finite. The model does not include a period of
rapid inflation, yet it
reproduces all of the successful predictions of
standard big bang and inflationary cosmology.
We point out numerous novel elements that have
not been used previously which may open the door to
further alternative cosmologies.
Although the model is motivated by M-theory, branes and extra-dimensions,
here we show that the
scenario can be   described almost entirely
in terms  of conventional 4d field theory and 4d cosmology.
\vspace{1pc}
\end{abstract}

\maketitle

\section{Introduction}
Through a combination of inspired theoretical
insights and ingenious experiments  and observations, cosmologists have
emerged with a consensus model 
model of the Universe, a mixture of the big bang picture and inflationary 
cosmology\cite{Gut,Lin}, 
that is able to explain a whole host of observations in exquisite 
detail. The results raise our confidence that we are converging on the right story of the evolution of the Universe.\cite{rev}  But maybe we're not!

The purpose of this paper is to introduce a new
 type of cosmological model -- 
not a variant of an older model, but, a genuinely new cosmological framework.
 This new paradigm turns cosmic history topsy-turvy.  
And yet, as you will see, it is able to reproduce all of the 
successful predictions of the consensus model with the same exquisite detail.
Perhaps you will even conclude that the cyclic Universe accomplishes the
feat more economically.

The key difference between the consensus paradigm  and the new paradigm 
can be simply stated.
The consensus model relies on the idea that space and time 
had a beginning when the 
Universe had nearly infinite temperature and density.   
It has been expanding every since, going from 
hot to cold, from dense to nearly vacuous.  The new model predicts 
that the Universe is infinite in both space and time. 
It is cyclic in time, undergoing endless cycles of evolution 
and renewal,  cooling and heating - 
in which the density and temperature remain finite throughout.  
All of the differences between the two paradigms harken back to 
the disparate assumptions about whether there is a  ``beginning''
or not.

The cyclic model of the Universe\cite{cyc1,cyc2}  
 draws heavily on ideas we developed earlier in 
 collaboration with Justin Khoury (Princeton), Burt Ovrut (Penn) and 
 Nathan Seiberg (IAS) in a precursor theory known as 
 the `ekpyrotic Universe.'\cite{kost,nonsing,ekperts}

Both cyclic and ekpyrotic scenarios
were very much inspired by string theory, M-theory and 
the notion of brane-worlds.  The introductory 
papers are expressed in this language.  However, as is  often the case, 
the notions that inspire an idea are not necessarily required.  
The idea may be more general.   So it is, we think, for the cyclic Universe.  
To make that point most forcefully, we will present this paper 
with hardly any mention of string theory until the very end. 
Instead, we will use the more prosaic and generic language of 
4d field theory, scalar fields, and potentials.  
Only at the end will we mention string theory  
 to show how it  provides a natural setting and simple
geometric interpretation of these ideas.

\section{The  Consensus Model}

To appreciate the cyclic model, it is useful to review the key
features and assumptions 
that underly  the consensus big bang/inflationary picture.

\vspace{.05in}
\noindent
{\bf Big Bang:}
The consensus model is based on the notion that the big bang is a beginning of 
space and time with nearly infinite
 temperature and density.  What is important is that the Universe 
starts out expanding, and there are patches over which
the inflaton field has high potential energy density. 
Regions of higher potential energy would expand faster, and
come to dominate. Hence one can argue that most of the 
Universe would be taken up by such high potential energy regions,
which would then inflate for a long time. 
This is the idea underlying chaotic inflation, for example.\cite{chaotic}
However, if 
the big bang were not a beginning, but, rather, a transition from a 
pre-existing contracting phase, then the inflationary mechanism
would fail (or
require major amendments).  During a
contracting phase, scalar field potential energy drives exponentially
rapid collapse
 so regions of high potential energy would shrink 
away faster than regions of low energy density (just the time reversed
behavior), so 
regions of high potential energy would not survive into 
the expanding phase to drive inflation.

The assumptions are all 
physically plausible and reasonable. However, we raise these points 
to emphasize that they are {\it unproven assumptions} that should not 
be taken for granted.  The theory is incomplete until they are proven.
The issue is important because,
as we shall see, the cyclic model makes different plausible
and reasonable assumptions about 
the nature of the bang.  

\vspace{.05in}
\noindent
{\bf Inflation:}
The consensus model assumes that, shortly after the big bang,
the Universe underwent a brief 
period of rapid, superluminal expansion: inflation. 
To have inflation, new ingredients have to be added to the simple
big bang picture. In typical models of inflation, the ingredients consist of 
a scalar field (the inflaton) and a scalar potential with the
property that, over some range of values, the field rolls so slowly
down the potential that its kinetic energy is much less than 
its potential energy. Figure~1(a) is a schematic illustration.
In this case, the slow-roll evolution occurs on the plateau (stage
1 in Figure 1(a)).
When the potential energy dominates, the pressure of the 
scalar field  (the difference between kinetic and 
potential energy density)
is negative, causing the Universe to undergo a 
period of cosmic acceleration. The height of the plateau is chosen
so that the acceleration is extraordinarily rapid, causing the 
Universe to double in size every $10^{-35}$~seconds (a fiducial number,
assuming GUT-scale inflation).  

\begin{figure}
\begin{center}
 \epsfxsize=3.25 in \centerline{\epsfbox{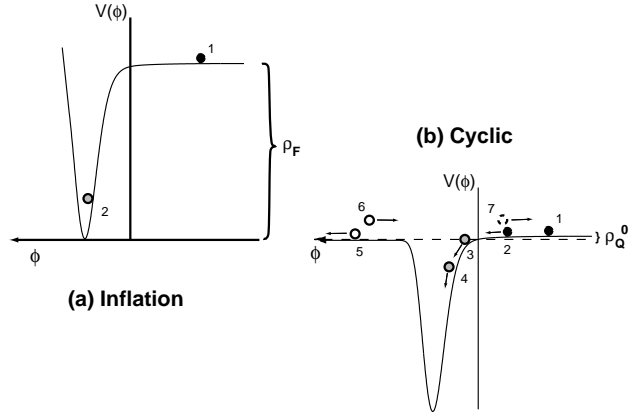}}
   \end{center}
       \caption{
       Schematic plot of the potential $V(\phi)$ as a function of the
       field $\phi$ for (a) inflationary cosmology and (b)
       cyclic models.  
	     The numbered sequence of
	    stages is described in the article. }  
				\end{figure}

It is essential that the slow-roll comes to an end in order to 
stop inflation. In the schematic example, the end of inflation
occurs where the scalar reaches the end of the plateau and falls
into the potential well (stage 2).  
The field oscillates around the minimum 
and decays into ordinary matter and radiation.

Despite the period of inflation which causes
the Universe to become homogeneous and flat, quantum fluctuations cause
different regions to reach the end of the plateau and to
reheat and cool  at different times,
 leading to fluctuations in the temperature and density.
The result of a careful analysis is that the spectrum of fluctuations
is nearly scale-invariant, gaussian and adiabatic,\cite{BST} in excellent 
accord with what is observed in measurements of the cosmic 
microwave background anisotropy and large-scale structure.  All this must be accomplished before the Universe is a second old in order 
to recover the 
successful predictions of primordial nucleosynthesis.  

\vspace{.05in}
\noindent
{\bf Radiation and Matter Domination:} When the Universe reheats
after inflation, the temperature is higher than the characteristic scales
of cosmic nucleosynthesis and structure formation, so the subsequent
expansion and cooling precisely mimics the original big bang picture
and all of the associated successful predictions  are re-accrued.
Nucleosynthesis occurs during the radiation epoch. The Universe
becomes transparent and structure forms in the matter-dominated epoch.

\vspace{.05in}
\noindent
{\bf Dark energy??:}  
Recently, the consensus model has had to incorporate a major 
amendment, the addition of dark energy. The  observations 
of the cosmic microwave background, large-scale structure and 
distant supernovae all suggest that most of the energy density of
the universe is some form of dark energy component with negative 
pressure which causes the expansion of the Universe
 to accelerate.\cite{supernova,rev} 

{\it The discovery of dark energy is a complete surprise from the point-of-view
of big bang and inflationary cosmology.} It serves no known
role.
It may consist of a vacuum energy (or cosmological
constant) or quintessence.   Although dark energy was not predicted,
it can be accommodated by stipulating
that the inflaton decayed into some combination of 
matter, radiation  and dark energy. 
However, the added conditions are {\it ad hoc},
and extreme 
tuning is required to explain the  ratio of dark energy to matter
energy.

\vspace{.05in}
\noindent
{\bf The future???:}  Because dark energy comes as a surprise and its
nature is not predicted, the long-term future of the Universe is
uncertain.  If the dark energy is a cosmological constant, the 
accelerated expansion will continue forever, and the Universe will
become increasingly empty.  If the dark energy is quintessence, 
many alternatives are possible. 
For example, 
 quintessence can decay and cause  the acceleration to 
end.

\section{The Cyclic Model}

Cosmic evolution in the cyclic model differs remarkably from the consensus picture:  

\vspace{.05in}
\noindent
{\bf The ``bang'':}  Each cycle can be said to begin with a bang.
Unlike  the traditional concept of the big bang, the density and
temperature of the Universe do not diverge, and, in the M-theory picture,
space does not disappear.  Instead, the bang is 
a transition or bounce from a pre-existing contracting phase
to an expanding phase
during which 
matter and radiation are created at a large but finite temperature.

\vspace{.05in}
\noindent
{\bf Radiation and Matter Domination:}
The bang is followed by an immediate entry into the radiation-dominated
epoch.  There is no inflation and no need for the scalar field 
or potential required by inflation.  Other mechanisms, to be discussed
below, are responsible for making the Universe homogeneous and flat and for
making density fluctuations.  	The radiation and matter-dominated
epoch are essentially identical to the consensus model.

\vspace{.05in}
\noindent
{\bf Dark energy:}  The Universe subsequently  enters the 
dark energy dominated epoch and cosmic acceleration commences.
Recall that dark energy was unanticipated in the consensus model and 
has no purpose. 
Here, 
dark energy plays a pivotal, and absolutely crucial role.  The field that 
comprises the dark energy is the engine that drives the whole 
cyclic scenario.  

The dark energy is due to a scalar field rolling down a 
potential, similar to the inflaton of the consensus model. 
However, here the scalar field governs the whole of the evolution 
of the Universe, not just the beginning.
 In inflation, the scalar field causes acceleration and reheating. 
 Here, the scalar field acts as dark energy, causes a
 period of slow acceleration, converts the acceleration 
into deceleration and contraction,
 triggers the bounce, reheats the Universe and begins the cycle anew. 
 Furthermore, the scalar field has a natural 
geometric interpretation in string theory,
as described in Section 7.

Compared to inflation, the potential is shifted
downwards in energy, as illustrated in Figure 1(b).
The plateau corresponds to a tiny potential
energy density equal to the current 
dark energy density. The scalar field acts as a form of 
quintessence, and so we have labeled the height of the potential
as $\rho_Q^0$. During the radiation and matter dominated phases,
the scalar field is effectively frozen in place by the Hubble
red shift of its kinetic energy (stage 1 in Fig.~1(b)).
However, as the Universe cools and expands, the potential energy 
ultimately comes to dominate that energy density.  Also, the field
begins to roll downhill (stage 2).  

The period in which the potential energy dominates is important for
several reasons.  It provides the source for the presently 
observed dark energy and cosmic
acceleration. It makes the Universe flat and homogeneous, replacing 
inflation.   Although both dark energy and inflation entail cosmic 
acceleration, the acceleration due to dark energy is 100 orders of
magnitude smaller, causing the Universe to double in size every 
15 billion years or so, compared to every
$10^{-35}$~seconds for inflation.
The very slow acceleration is, nevertheless, sufficient to
empty the Universe of its matter and radiation.
After trillions of years, 
there is
less than one particle per horizon.  Locally, the Universe has
been restored  to nearly pristine vacuum.   

The emptying of the Universe is crucial to a cyclic model.  Historically,
oscillatory models have been plagued by the fact that the entropy density
rises from cycle to cycle, as shown by Richard Tolman in the 
1930s.\cite{Tolman}   The lengths of cycles increase steadily.
Consequently, extrapolating backwards in time, the lengths of cycles
decrease steadily, so that the sum converges at a finite time
-- a beginning of the Universe that one was trying to avoid.
In our cyclic model, the dark energy dilutes the entropy density 
to negligible levels at the end of
 each cycle, preparing the way for a new cycle of identical 
duration.  

Note that the second law of thermodynamics is respected. The total
entropy of the Universe rises from cycle to cycle. The number of 
black holes rises as well. (Equivalently, the entropy and black holes
per unit comoving volume increase.)  However, the physical entropy density
 -- the entropy per proper volume -- is expanded away in each cycle.
Since it is the physical entropy density which determines
 the expansion rate, the expansion and contraction history 
in each cycle is the same from one cycle to the next.
(A key feature is that
 entropy density decreases in the expansion phase but, as
we shall see, does not grow significantly during the contraction phase due
to the effects of the scalar field, countering the effects of contraction.)

It may seem peculiar that the entropy density remains finite at the
crunch.  That is, even if the entropy density is made exponentially
small during the dark energy dominated phase, why doesn't it diverge?
The answer will be that the crunch is modified by the interaction
between matter-radiation and the scalar field.

\vspace{.05in}
\noindent
{\bf Contraction:} As the field rolls off the plateau and 
heads downhill, the potential energy crosses zero (stage 3).
The Universe is dominated by the kinetic energy of the scalar
field, which causes the expansion to decelerate once again. 
As the field rolls to values where the potential energy is
increasingly negative, the expansion stops altogether (stage 4) and 
reverses to contraction.  

The contraction is extremely slow at first, taking billions of 
years. During this very slow evolution phase, quantum fluctuations
have time to cause spatial variations in the
rate of contraction.
At first, these variations are only on large length scales, but, as
time proceeds, they occur on smaller and smaller length scales.
The big surprise, as first  demonstrated by Khoury {\it et al.} in 
the context of the ekpyrotic model,\cite{kost,ekperts}
is that the resulting spectrum 
is scale-invariant if the potential flattens exponentially at large $\phi$. 
(Small deviations from exponential behaviour
result in small deviations from
scale-invariance.)

Differences in the contraction rate result in differences in when 
different regions bounce, reheat to high temperature and expand. 
Consequently, the scale-invariant spectrum generated in the 
contraction phase evolves to become a spectrum of temperature and density
perturbations after the bang.
(The detailed computation of perturbations in the 
ekpyrotic and cyclic models led to fierce debate in the literature
over the precise matching conditions. Matching conditions
according to which the scale-invariant spectrum generated in the 
contraction phase propagate across the singularity have been proposed,
and whilst these have not yet been fully justified at a fundamental
level, progress has been made 
\cite{durrer,tolley,moore}.)

\vspace{.05in}
\noindent
{\bf ``Crunch" and ``bounce":}  The scalar field picks up speed
as it rolls down into the potential well.  One need not fear that
the field will be stuck at the bottom. Contraction causes a Hubble
blue-shift of the scalar field kinetic energy.  The field literally 
accelerates out of the minimum and flies off towards negative 
infinity in a finite time (stage 5).   At the same time, the scale 
factor in the effective 4d field theory rushes towards zero and
the crunch ensues. 

Unlike a conventional big crunch, the temperature and density remain
small and finite before the crunch.  This is because the scalar
field is coupled to  matter and radiation in a special way. 
The effect of contraction in increasing the density is directly
compensated by the interaction with the scalar field that drains
the energy density.  We will illustrate how this is possible and 
well-motivated in M-theory.

The field reaches negative infinity in finite time, before
heading back towards
positive values (stage 6). The acceleration of the field at turnaround
causes some conversion of scalar field kinetic energy to matter and
radiation, reheating the Universe to high but finite temperature.
At the same time, the scale factor reverses and the Universe 
begins to expand.  The field rushes back towards  where it started
(stage 7), but
its kinetic energy is now being red-shifted in the expanding phase,
especially due to the presence of matter and radiation. Soon after
the radiation dominates over the scalar kinetic energy density, 
the red-shift causes the field to grind to a halt (stage 1).

The Universe has now come full circle, returning
to the expanding, radiation-dominated phase.

\section{Another view: Expansion-Stagnation}

We have described the cycle as a sequence of expansion
contraction.  Technically, we call the second stage contraction
because the scale factor $a(t)$ is shrinking and, at the crunch,
is equal to zero.  However, as pointed out above, the scalar 
field conspires in such a way as to keep the temperature and density
nearly fixed at a small value.  In the brane picture, the expansion
corresponds to the stretching of the branes, which dilutes and
matter and radiatoin density.  The contraction, though, is the 
contraction of the finite extra dimension, rather than the three large 
dimensions. 

Consequently, an ordinary observer on our brane, say,
would not recognize the behavior at the end of  trillions of years of
expansion
 as a contraction
since the interval between distant objects would not appear 
to be shrinking in brane coordinates.
Rather, since the branes stop stretching,
it would appear that the expansion has stalled. The contraction of 
the extra dimension would not visible, but its physical consequences
would be.  Namely, coupling constants would change with time at
a rate that increases as the bounce approaches. This would be perceived
as some mysterious new form of energy.  At the bounce, 
this energy would suddenly produce radiation and energy that fills the
universe.  As the branes bounce back, the couplings are  returned to 
their initial values.  The matter and radiation force a new period of
expansion (brane-stretching).   

In sum, if the story is retold from the ordinary observer's 
point-of-view, the universe is going through cycles of expansion,
stagnation, reignition and then the cycle begins again.

\section {Consensus vs. Cyclic}

We have outlined the basic elements of two cosmological 
scenarios based on different assumptions about the nature of 
space and time. The consensus picture assumes that space and time have
a beginning at the big bang.  The cyclic model assumes that 
the bang is simply a transition from contraction to expansion
accompanied by the release of radiation.  

Figure 2 compares the two scenarios. Several points are apparent.
\begin{verse}
{\it Inflation and Cyclic models entail two completely different
strategies for making the Universe homogeneous and flat and for
generating density perturbations, each strategy resting on the respective
assumptions about the beginning of the Universe.}
\end{verse}
For Inflation, there is
little time before primordial nucleosynthesis 
to fix conditions on large scales. An ultra-rapid period of accelerated
expansion is called for.  The cyclic model assumes that time continues
before the bang, so there is plenty of time to set the large-scale conditions
of the Universe prior to the bang.  Hence, it fits quite naturally
to invoke long periods  of ultra-slow
accelerated expansion and contraction lasting trillions
of years during which the homogeneity, 
flatness, and density perturbations are established.
For comparison, inflationary fluctuations are amplified and
frozen in on a timescale of $10^{-35}$ seconds, a length scale
of $10^{-25}$ centimetres. Whereas fluctuations are generated
fractions of a second before the big bang, when their length
scale is thousands of kilometres.

\begin{figure}
\begin{center}
 \epsfxsize=3. in \centerline{\epsfbox{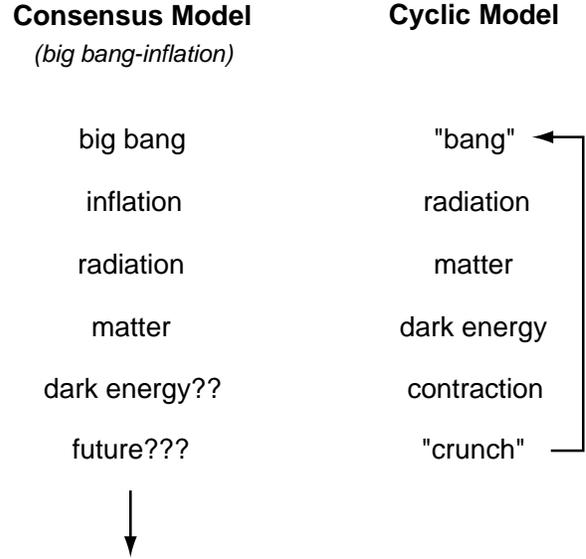}}
   \end{center}
       \caption{
	    Comparison of the sequence of events in the
	    consensus big bang/inflationary and cyclic pictures.
	    The consensus model assumes that the big bang is a
	    beginning, inflation sets the initial conditions, 
	    and then the Universe proceeds through the 
	    matter and radiation dominated phases. 
	    Dark energy is added to explain the current acceleration,
	    but the long-term future is uncertain.  The cyclic model
	    proposes that the Universe undergoes an endless sequences
	    of cycles which begin with a bang and end in crunch.
	    Initial conditions for a given cycle are produced during the
	    dark energy-dominated and contraction phases of the 
	    previous cycle.  The past and future are determined.}
				\end{figure}

\begin{verse}
{\it Inflation requires two periods of acceleration with two different
energy components, whereas the cyclic model requires only one. }
\end{verse}
For inflation, we must introduce a hypothetical period of 
accelerated expansion
driven by an inflaton field and potential with certain properties. 
One must  add a second component to explain the observed 
acceleration today.  Although one can imagine both accelerations being
due to the same field, there seems to be nothing to recommend it: the 
time and energy scales are so different that there is no natural link.
By contrast, the cyclic model requires only one period of accelerated
expansion, the one that is actually observed.
Some corollaries are:

\begin{verse}
{\it Dark energy is not predicted or explained by Inflation, but
is a required component in the Cyclic Model.}
\end{verse}

\begin{verse}
{\it The future is not predicted by Inflation, but it is predicted in the
Cyclic Model.}
\end{verse}

More generally, it is obvious that Inflation is  a theory of the 
very early Universe, without explanation of the beginning or future of the
universe. By contrast,
 the cyclic model offers a complete history.  Knowledge of the present is 
 also knowledge of the past and future. Consequently, it is not surprising
 that the cyclic model is more constrained.  
 For example, the amplitude of long-wavelength gravitational waves
 is constrained to be
 exponentially tiny in the cyclic model whereas
 there is freedom to adjust the amplitude in Inflation.
 Nevertheless, all current observational constraints are satisfied
 by both models.

 That the cyclic model is able to accomplish more with less suggests that
 it is economical.  However, economy is not a reliable way to judge
 between the two scenarios.  The critical issues are bith theoretical -- the 
 nature of the bang -- and experimental - searching for  long-wavelength gravitational
 waves.

\section{Contraction and Generation of Perturbations}

The cyclic model introduces several novel features into cosmology.
In this introduction, we will focus on the  two  most important:
the new mechanism for generating a scale invariant spectrum of density
perturbations and the bounce from crunch to bang.

An important discovery by Khoury {\it et al.} in our development of the 
ekpyrotic model was a new mechanism for generating a scale-invariant
spectrum of density perturbations using a phase of slow contraction rather
than rapid inflation.\cite{kost,ekperts}
The fact that both approaches work is shocking, at first, because
they seem so different.  Gravity is the strong, overwhelming influence
during inflation, but it is essentially
irrelevant during  slow contraction.

Actually, there is a simple heuristic argument to explain the result. During
slow contraction, gravity can be ignored in
the equation of motion for the scalar field $\phi$. Its integral form is
simply:
\begin{equation}
\dot{\phi} = \sqrt{-2 V}.
\end{equation}

For Inflation, the idealized model is a perfectly flat potential.
For cyclic models, the analog is a pure, decreasing exponential potential,
\begin{equation}
V= -V_0 {\rm exp} (-\sqrt{3 (1+w)} \phi),
\end{equation}
where this form is taken to be an approximation to the steeply falling
part of the potential after the plateau.  The solution to the full equations of motion in a gravitational background for an exponential potential is well-known.
There is a scaling solution in which the field energy is that of an ideal fluid with equation of state $w$, 
in units where $8 \pi G=1$.  Note that the same potential form can be used for inflation if $V$ is negative and $w \approx -1$. In cyclic models, we need $V_0$ positive and 
$w \gg 1$. 

The equation of motion can then be integrated to provide a relation
between the time and the curvature of the potential:
\begin{equation}
-t = \sqrt{\frac{2}{-V''}},
\end{equation}
where the time has been defined to be negative and to approach $t=0$
as the contraction reaches the big crunch.
The equation of motion for the $k$-th  fluctuation mode, $\delta \phi_k$
is, then:
\begin{equation}
\delta \ddot{\phi}_k +  (k^2 + V'') \delta \phi_k =  0
\end{equation}
the last term of which can be re-expressed in terms of the time
using the solution to the equation of motion:
\begin{equation} 
\delta \ddot{\phi}_k +  \left(k^2 - \frac{2}{t^2} \right) \delta \phi_k =  0.
\end{equation}

Those familiar with inflationary perturbations will recognize the last
equation as being precisely what is obtained in inflation for 
a flat potential.   The critical factor is the numerator of the second
term, which is two for a precisely scale-invariant spectrum. 
The factor of $2/t^2$ for inflation comes from the gravitational 
expansion  term, which is proportional to $a''/a = 2/ \tau^2$ where 
$a$ is the scale factor and 
the prime represents the derivative with respect to conformal time $\tau$.
The  magical result is that scale invariance can be 
obtained either when gravity is strong and expansion is nearly de Sitter, 
or when contraction is negligible and the potential is 
exponentially decreasing. 

However, there is an important difference. In inflation, the 
scale-invariant perturbations are generated because the
gravitational factor  $a''/a$ is 
 much greater than the mass
of the inflaton $V''$.  
A corollary is that
the same equation applies to any field whose mass is 
much less than $a''/a$.   That is, all light degrees of freedom obtain
scale-invariant fluctuations. For most degrees of freedom, these 
fluctuations are irrelevant because they either smooth out
after re-thermalization, or 
never dominate or never leave a  measurable imprint.  Notable
exceptions are the two massless tensor modes.
The gravitational waves produced in inflation may not dominate
the density of the Universe, but they decouple immediately after
reheating and maintain a distinctive signature.
Although the current generation of gravitational wave
detectors is not sensitive enough,
future detectors  may detect directly the stochastic fluctuations
created by inflation. Alternatively, 
the tensor fluctuations leave a distinctive
signature in the cosmic microwave background polarization, which also
may be detectable.   The discovery of long-wavelength gravitational waves
would be important proof for
inflation because these long-wavelength gravitational waves are not
produced in the cyclic or ekpyrotic models. For cyclic models, 
the critical factor of $2/t^2$ is provided by the potential, so 
only fields which have the exponential potential obtain scale-invariant
fluctuations.  The components of the metric are massless and
potential-less, so, to lowest order, 
there are no long-wavelength gravitational
waves produced.  (When gravitational back-reaction is included,
one finds a blue spectrum which includes 
long-wavelength modes, but whose amplitude is $10^{25}$ smaller than 
inflation -- effectively invisible.\cite{kost})

Hence, a valuable lesson has been learned. Gravitational waves provide
a litmus test for determining the conditions under which the primordial 
perturbations were created, during fast expansion or slow contraction.

Another difference, even harder to detect, is the higher order 
non-gaussian contributions to the primordial spectrum.  In both models,
one can compute the non-gaussian corrections to the leading order gaussian
fluctuations. The inflationary contribution is very small, down by 
more than $10^{-5}$ relative to the leading order.   Cyclic perturbation
spectra are super-gaussian: the non-gaussian contribution due to higher
order terms is exponentially suppressed.\cite{future}

\section{Contraction, Crunch and Bounce}

If the Universe is contracting when the perturbations are created, 
it will continue to contract until the Friedmann-Robertson-Walker
scale factor, $a(t)$, hits  zero and the Universe undergoes a crunch.
The proposal is that the crunch leads to a bounce and renewed
expansion.

One of the lessons learned from the study of the ekpyrotic 
model is that the crunch may be much milder than expected.\cite{nonsing} 
The big crunch discussed in earlier attempts at oscillatory models
was one in which the density and temperature diverge, leading 
to divergent curvature and uncontrolled behavior.  
For example, the action might be:
\begin{equation}
S = \int d^4x a^4  \left(  \frac{1}{2} {\cal R} +
\frac{1}{2} (\partial \phi)^2  - 
V(\phi) + \rho_R \right)
\end{equation}
where $8 \pi G=1$, ${\cal R}$ is the Ricci scalar and 
$\rho_R$ is the radiation density.  We have also set
$\sqrt{-g}= a^4$,  using the conformal Friedmann-Robertson-Walker
metric.
Solving the equation of motions, one finds
$\rho_R \propto a^{-4}$, which is divergent at the crunch.

However, the crunch can be mollified if the scalar field $\phi$
has appropriate couplings to matter, as in:
\begin{equation}
S = \int d^4x a^4  \left(  \frac{1}{2} {\cal R} +
\frac{1}{2} (\partial \phi)^2  -
V(\phi) +  \beta^4(\phi) \, \rho_R
\right).
\end{equation}
If the coupling function $\beta(\phi)$ has the property that 
$a \, \beta \rightarrow \, cnst.$ as $a \rightarrow 0$, then 
 the radiation density approaches a constant at the crunch
because the solution to the equation of motion is modified 
to be $\rho_R \propto (a \, \beta)^{-4}$. 
The increase in density due to contraction  is compensated by the 
effect of the scalar field, which acts as a modification of gravity.
By this simple 
mechanism, some of the serious concerns are obviated.
The density and curvature can be finite, at this classical level.
The spatial  singularity has properties akin to time-like conical
singularities which have been resolved and tamed in string theory.
	Our speculation 
is that the cosmic singularity  can be tamed, also, even
after the inclusion of quantum fluctuations.  Efforts are underway 
to evaluate this speculation rigorously.  

Effectively, this approach argues that the effect of the contraction
of the scale factor on the density curvature is compensated by the 
effect of the scalar field, which acts as a modification of gravity.

\section{A New Toolkit for the Cosmologist}

We have emphasized that the cyclic scenario can be understood 
for the most part, in terms of
ordinary field theory.
Many of its  basic ingredients, such as scalar fields and potentials, 
also appear
in inflationary cosmology.   The cyclic model  could have been 
proposed at the same time  as inflation, in which case it would
have been interesting to see which proposal would have emerged as the more 
appealing idea. 
As it is, the cyclic model has arrived twenty years late, after cosmologists have become
attached to the inflationary paradigm, so there can be no fair measure.   

The cyclic model was probably delayed  because of 
the number of novel concepts that had to be incorporated.  Although 
each idea
is rather simple, 
having the whole collection together at once  opens the door
to new cosmological possibilities. 
They form a new toolkit of 
ideas that enable the cyclic model and 
may inspire further alternative cosmologies. For this reason, we
summarize them here:

\begin{itemize}
\item Negative potential energy can be used 
to trigger  a period of collapse in a flat universe. 
(Previous oscillatory models assumed a closed universe.)

\item
Negative potentials  can lead to viable
	cosmologies.  
	
	(Many thought that the field would roll down
	the potential, get stuck at the bottom, and collapse into
	an anti-deSitter phase. Instead,  the 
	collapse causes the kinetic energy of the scalar field to
	blue shift and accelerate out of the potential well.
	This result has important implications for model-building.
	For example, supergravity models have been summarily  rejected in the
	past because the vacuum state had negative potential 
	energy, but now we have learned that some of the models
	can lead to viable cosmologies.\cite{banks})

\item 
A period of dark energy domination can  simultaneously resolve the 
	entropy problem of cyclic models, and the homogeneity
	and flatness problems of the standard big bang model.

\item 
 Scale-invariant spectra can be generated during a slow
	contraction phase in which $w$ for the scalar field
	is much greater than unity, as explained in Section~4.

 \item  Scalar field couplings to matter and radiation 
 can be chosen so that the
 (classical) density and temperature 
	remain   finite at collapse. 

\item
Density fluctuations created in a collapsing phase can
	result in growing mode perturbations after a bounce.\cite{ekperts}

\item 
Dark energy can  make the cyclic solution a dynamical
	attractor so that the
	solution that is stable  under perturbations.
\end{itemize}

\section{Strings and Branes}

Thus far, the discussion of 
strings and branes has been avoided. 
However, string theory and M-theory have been an  inspiration because
they provide 
 a natural setting and a particularly appealing geometric
realization of the cyclic scenario.
This connection is important, giving the model 
quantum consistency at a deep level and connecting
our scenario to the leading approach to
fundamental physics. 

According to $M$-theory,
the Universe
 consists of a four dimensional
 `bulk' space bounded by two three-dimensional  domain walls,
 one with positive and the other with negative
 tension.\cite{REFBRANE,HW,polch} The
 branes are free to move along the extra spatial
 dimension, so that they may approach and collide.
 (The fundamental theory  is formulated in 10 spatial dimensions, but
 six of the dimensions are compactified on a Calabi-Yau
 manifold, which for our purposes can be treated as fixed and 
 irrelevant).
 Gravity acts throughout the five dimensional space-time, but
 particles of our visible Universe are constrained
 to move along one of the branes, sometimes called the
 `visible brane.'   Particles on the other brane interact
 only through gravity with matter on the visible brane and hence behave like
 dark matter.

Now we can ``name names."
The scalar field $\phi$  responsible for the cyclic scenario is, here,
naturally identified with the radion
field that determines the distance between branes. It is a required 
component of the M-theory picture.   The potential
$V(\phi)$ is the inter-brane potential caused non-perturbative
virtual exchange of membranes between the boundaries.  
The interbrane force is what causes the branes to repeatedly
collide and bounce.  
The particular form of the interbrane force is not known, but 
a plausible form for the interbrane  potential is:
\begin{equation}
V(\phi) = V_0 (1- e^{-c\phi}) F(\phi) 
\end{equation}
where $F(\phi)$ is chosen so that the potential falls to 
a substantially negative value and than rapidly approaches zero
as $\phi \rightarrow -\infty$. 
At large separation (corresponding to large, positive $\phi$), 
the force between the branes should become small, consistent with
the flat plateau shown in Fig.~1(b).  Collision 
corresponds to $\phi \rightarrow -\infty$. 
But the string coupling 
$g_s \propto e^{\gamma \phi}$,  with $\gamma > 0$, so $g_s$ approaches 
zero in this limit.\cite{nonsing} Non-perturbative effects vanish
faster than any power of $g_s$, for example as 
$e^{-1/g_s^2} $   or $e^{-1/g_s}$, accounting for the
prefactor $F(\phi)$.

The coupling to matter-radiation, $\beta(\phi)$,  
that mollifies the crunch (see Section~5)
also
has a natural interpretation in the brane picture.  
Particles reside on the branes, which are embedded in an
extra dimension whose size and warp are determined by 
$\beta$.  The effective scale factor on the branes is
$\hat{a} = a \, \beta(\phi)$, not $a$, and
$\hat{a}$ is finite at the big crunch/big bang.  The function 
$\beta(\phi)$ is, in general, different for the two branes (due to the 
warp factor) and for different reductions of M-theory.  However,
the behavior 
$\beta(\phi) \sim {\rm e}^{-\phi/\sqrt{6}}$
as $\phi \rightarrow - \infty$ is universal, since at small
brane separations the warp factor becomes irrelevant and one
obtains the 
standard Kaluza-Klein result.\cite{nonsing,cyc1}
This universal form satisfies the desired condition that $a \, \beta
\rightarrow const.$ as $a \rightarrow 0$.

Most importantly, the brane-world provides a natural resolution
of the cosmic singularity.\cite{nonsing}
From the brane-world perspective,
the singularity   is far milder than in conventional
cosmology.   In fact, one might say
the big crunch is an illusion, since the scale factors
on the branes ($\rightarrow \hat{a}$)  are perfectly finite there. 
That is why the matter and radiation densities, 
and the Riemannian curvature on the branes, are finite.  
The only respect in which the 
big crunch is singular is that 
the one extra dimension separating the two branes 
momentarily disappears. 
{\it That is, the apparent singularity of the 4d theory, $a \rightarrow 0$,  
 corresponds to the collapse of
the extra dimension only.  Our regular
three dimensions  remain infinite.}

Our scenario is built on the hypothesis\cite{kost} that 
the branes separate after collision, so the extra dimension
immediately reappears.
Consequently, the scale factor bounces and begins to expand.
This process cannot be completely
smooth, since the disappearance of the extra dimension
is non-adiabatic and leads to particle production. Preliminary
calculations of this effect are encouraging, since they
indicate a finite density of particles is produced\cite{newpolch,tolley}.
In this picture,  
the brane collision can be viewed as a simple, partially inelastic collision.
Ultimately, a well-controlled string-theoretic 
calculation\cite{nonsing,tolley,moore}
should determine the efficiency of particle production from 
first principles.

in this picture, an observer on a brane would experience
periods when the brane is expanding (during the radiation, matter,
and dark energy dominated epochs), and periods when the stretching 
of the brane would nearly halt.   instead, the branes approach
one another. The observer would perceive that the separation between
distance objects on the brane 
is no long increasing, nor is it decreasing either.
This is the stagnation discussed earlier. Rather, the action is 
occuring in the extra dimension.  Once the collision between branes
takes place and the universe is dominated once again by radiation,
the expansion of the universe begins again.  
  
 On length scales larger than the separation between
 branes,  the higher
 dimensional brane-world description can be reduced to 
 an effective  four-dimensional field theory.  This is why, for
 the most part,
  the cyclic scenario could  be described without reference to
 stringy physics and higher dimensions.  
   Nevertheless, 
 the brane-world picture proves to be a useful geometrical
 picture,  and string theory is probably essential
 for determining what happens at the crunch.

\section{In the Beginning}

With the 
introduction of the cyclic model,
the stage is set for a scientific debate  comparing  two distinct 
conceptual frameworks,
one  based on the supposition that space
and time have a beginning  and the other  based on the notion that 
time and space continue and 
 the bang is merely a transitional stage
from one cycle to the next.  Conceptually, the two models are poles apart.
Observationally, both match all current observations in exquisite 
detail.  

To decide  between the two models  
requires advances in both theory and experiment. Theory, particularly
string theory, can shed light on the nature of the singularity.  As 
a candidate for a quantum theory of gravity, addressing the singularity is
its {\it raison d'etre}.  The fact that the singularity in question
is mathematically equivalent to a collision between branes with finite 
energy density in a non-singular
region of 5d space-time strongly suggests to these authors that time does not
come to a halt at the crunch.  Intuitively it seems that the branes must bounce
(or equivalently, 
pass through one another), re-opening the extra dimension, corresponding to
reversal from contraction to expansion.  However, establishing
this is the critical
challenge for string theory and  must be shown 
rigorously.
  String theory can also
inform us whether the brane interaction
can be of the requisite form for the model.  

The ultimate arbiter will be Nature. Specifically, measurements
of the stochastic gravitational background is the decisive way
to distinguish the two scenarios.
Another generic
prediction of the cyclic model regards the ratio of the pressure to
the energy density of the dark energy that is causing the current
cosmic acceleration.  In the cyclic picture, the dark energy is due
to the scalar field $\phi$
which has been fixed during the radiation and
matter era, but is beginning to roll downhill as the Universe
becomes dark energy dominated and the expansion begins to accelerate.
For a static field, the ratio of pressure to energy density is -1,
but this ratio increases as the field begins to roll.
Hence, measurements of  the ratio today and perhaps its time-variation are
further consistency checks of the cyclic picture.
In the interim, it appears that we
now have two disparate possibilities: a Universe with a definite
beginning and a Universe that is made and remade forever.

\section{Acknowledgements}
We would like to acknowledge our valued collaborators
Justin Khoury, Burt Ovrut and Nathan Seiberg.
    The work was supported in part by
      US Department of Energy grant
      DE-FG02-91ER40671.

\end{document}